\documentstyle[11pt]{article}
\newcommand{\s}{\sigma }                                              
\newcommand{\SI}{\Sigma }    
\newcommand{\tS}{\tilde{\Sigma}}
\newcommand{\tG}{\tilde G}    
\newcommand{\si}{\sigma _1}  
\newcommand{\st}{\sigma _2}

\newcommand{\xn}{x_{n}}
\newcommand{\xns}{x_{n}(\sigma )}
\newcommand{\xnsp}{x_{n}(\sigma ')}
\newcommand{\xm}{x_{m}}
\newcommand{\xms}{x_{m}(\sigma )}
\newcommand{\xmsp}{x_{m}(\sigma ')} 
\newcommand{\e}{e^{i\int k_{0}Y}}                                      
\newcommand{\qe}{e^{iq_{0}Y}}
\newcommand{\kim}{ k_{1}^{\mu}}                                      
\newcommand{\kom}{ k_{0}^{\mu}}                                      
\newcommand{\ki}{ k_{1}}
\newcommand{\yn}{ Y_{n}}                                             
\newcommand{\ym}{ Y_{m}} 
\newcommand{\kn}{ k_{n}}
\newcommand{\Kn}{ K_{n}}
\newcommand{\km}{ k_{m}}
\newcommand{\Km}{ K_{m}}
\newcommand{\kt}{ k_{2}}                                             
\newcommand{\ko}{ k_{0}}                                             
\newcommand{\yim}{ Y_{1}^{\mu}}                                      
                                      
\newcommand{\kin}{ k_{1}^{\nu}}                                      
\newcommand{\kon}{ k_{0}^{\nu}}                                      
\newcommand{\ktm}{ k_{2}^{\nu}}                                      
\newcommand{\ytm}{ Y_{2}^{\mu}}                                      
                                                              
\newcommand{\lpp}{\mbox {$e^{i\int _{c} \alpha (t)                             
k(t) \partial _{z} X(z+t) dt +ik_{0}X}$}}

\newcommand{\gvk}{ e^{i\sum _{n }k_{n}Y_{n}}}                      
                     
\newcommand{\dsi}{\frac{\partial}{\partial x_{1}}}             
\newcommand{\dsic}{\frac{\partial ^{3}}{\partial x_{1}^{3}}}         
\newcommand{\dsis}{\frac{\partial}{\partial x_{1}(\sigma _{1})}} 
\newcommand{\dsts}{\frac{\partial}{\partial x_{1}(\sigma _{2})}} 
\newcommand{\dstx}{\frac{\partial}{\partial x_{2}(\sigma _{2})}}
\newcommand{\dsix}{\frac{\partial}{\partial x_{2}(\sigma _{1})}} 
         
\newcommand{\dsn}{\frac{\partial }{\partial x_{n}}}             
\newcommand{\dsq}{\frac{\partial }{\partial x_{n+m}}}           
\newcommand{\dst}{\frac{\partial }{\partial x_{2}}}             
        
\newcommand{\dsth}{\frac{\partial }{\partial x_{3}}}

\newcommand{\dsit}{\frac{\partial ^{2}}
{\partial x_{1}\partial x_{2}}}
\newcommand{\dsb}{\frac{\partial ^{2}[\tS +\tilde G] (\si ,\st )}{\partial
 x_{1}(\si )\partial x _{1}(\st )}}
\newcommand{\dsnm}{\frac{\partial ^{2}}
{\partial x_{n}\partial x_{m}}}
\newcommand{\dsii}{\frac{\partial ^{2}}
{\partial x_{1}^{2}}}
\newcommand{\p}{\partial}                                           
\newcommand{\pp}{\partial ^{2}}                                     
\newcommand{\mup}{\partial _{\mu}}                                  
                                  
\newcommand{\eA}{\be    \label{lv}
e^{i\{ \int d\s\ko (\s ) Y (\s )+ i\sum _{n>0}  \kn (\s ) \frac{\p
Y(\s )}{\xns}
 \} }\]\[e^{ \int \int d\si
d\st \{ \ko (\si )\ko ( \st )[\tS +\tilde G](\si ,\st ) + (\sum _{n>0} \kn
(\si ) .
 \ko (\st ) \frac{\p
[\tS + \tilde G] (\si , \st )}{\p \xn (\si )}+\si \leftrightarrow \st) \}} \] 
\[e^{\int \int d\si d \st \{ \sum _{n,m>0}\kn (\si ) .\km (\st )
\dsb  \}}
\ee}
\newcommand{\li}{ \lambda_{1}}                                    
\newcommand{\lt}{ \lambda_{2}}                                    
\newcommand{\eps}{ \epsilon}                                        
\newcommand{\al}{\alpha }                                             
\newcommand{\aln}{\alpha _{n}} 
\newcommand{\tY}{\tilde Y}

\newcommand{\qin}{\mbox {$ q_{1}^{\nu}$}}

\newcommand{\ai}{\mbox{$\alpha _{1}$}}

\newcommand{\at}{\mbox{$\alpha _{2}$}}

\newcommand{\la}{\mbox{$ \lambda $}}                                           
\newcommand{\be}{\begin{equation}}                                             
\newcommand{\br}{\begin{eqnarray}}                                             
\newcommand{\ee}{\end{equation}}                                               
\newcommand{\er}{\end{eqnarray}}                                               
\newcommand{\eln}{\mbox {$ e^{\sum _{n}\lambda _{-n}L_{+n}}$}}

\renewcommand{\theequation}{\thesubsection.\arabic{equation}}

\begin{document}                                                               
\title{
\hfill\parbox{4cm}{\normalsize IMSC/2000/02/07\\
                               hep-th/0002139}\\        
\vspace{2cm}
Loop Variables and Gauge Invariant Interactions - I
\thanks{This is a detailed description
of an
 approach, outlined in a talk at the
Puri Workshop
in 1996, to use loop variables to string interactions.}}
\author{B. Sathiapalan\\ {\em                                                  
Institute of Mathematical Sciences}\\{\em Taramani                     
}\\{\em Chennai, India 600113}}                                     
\maketitle                                                                     

\begin{abstract}                                                               
We describe a method of writing down interacting 
equations for all the modes of the bosonic open string. It is
a generalization of the loop variable approach that was used
earlier for the free, and  lowest order interacting cases.
The generalization involves, as before, the introduction of a parameter
to label the different strings involved in an interaction.  The interacting
string has thus becomes a ``band'' of finite width. 
The interaction
equations
expressed in terms of loop variables, has a simple invariance that is
exact even off shell. A consistent definition of space-time fields
requires the fields to be functions of all the infinite number of 
gauge coordinates (in addition to space time coordinates). 
The theory is formulated in one higher dimension, where the modes
appear massless. The
dimensional reduction that is needed to make contact with string
theory (which has been discussed earlier for the free case) 
is not discussed here.

\end{abstract}                                                                 
\newpage                                                                       
\section{Introduction}                                                         
The loop variable approach introduced in \cite{BS1} (hereafter I)
(see also \cite{BS3})
is an attempt to write
down gauge invariant equations of motion for both massive and massless 
modes.  This method being rooted in the sigma model approach 
\cite{C,S,DS,FT,Polch,T},the  
computations are expected to be simpler and the gauge transformation
laws more transparent.  This hope was borne out at the free level
and also to a certain extent in the interacting case \cite{BS2} (
hereafter II).  The gauge transformations at the free level 
can be summarized
by the equation
\be \label{1.1}
k(t) \rightarrow k(t) \lambda (t)
\ee
Here $k(t)$ is the generalized momentum Fourier-conjugate to $X$ and
$\lambda$ is the gauge parameter.  This clearly has the form
of a rescaling and one can speculate on the space-time interpretation
of the string symmetries as has been done for instance in I. 

In II the interacting case was discussed.  It was shown that the leading
interactions could be obtained by the simple trick of
introducing an additional parameter `$\sigma $' as $k(t) \rightarrow
k(t, \sigma )$, parametrizing different interacting strings.
Thus, for instance, $\kim (\sigma _{1} )\kin (\sigma _{2})$ could stand
for two massless photons when $\sigma _{1} \neq \sigma _{2}$, but when
$\sigma _{1} = \sigma _{2}$ it would represent a massive ``spin 2''
excitation of one string.  The gauge transformations admit a corresponding
generalization
\be           \label{1.2}
k(t, \sigma )\rightarrow k(t,\sigma )\int d\sigma _{1} 
\lambda (t,\sigma _{1})                                  
\ee
It was shown, however that this prescription introduces only the
leading interaction terms.\footnote{Our notation, unfortunately, is
perverse: The variable $t$ originally used in the free theory lies
{\em along} the string, and $\s$ introduced in II labelling as it
does the number of interaction vertices, parametrizes {\em evolution}.}

In a talk some years ago \cite{Puri} (hereafter III)
 we showed that there is a natural generalization of this 
construction to include the full set of interactions that one
expects based on the operator product expansion (OPE) of
vertex operators.  
It was shown that this construction gives gauge invariant equations. 
The generalization involves introducing $\s $-dependence in the $X$
coordinates also. Gauge invariance at the level of loop variables is
very easy to see. What was not clear at the time was whether there was a
 consistent map to space-time fields. Here we show that this is in
 fact the case. It crucially involves keeping a finite cutoff on the
 world sheet and also making the space-time fields a function of
 $\xn$. Keeping a finite cutoff is required when going off shell
 \cite{BSPT,BSOS,AD}.
In the presence of a finite cutoff 
there  are problems with gauge invariance as discussed in
 \cite{BSFC}. It was shown there that to lowest order these problems
 could be resolved by adding a massive mode with an appropriate
 transformation law. It was also speculated that maintaining exact
 gauge invariance would be possible if all the modes are kept. This is
 shown to be true in the present work. We have the full gauge
 invariance and it is not violated by a finite cutoff and the
 construction necessarily requires all the modes.

Another feature that emerges from the present work is that the
space-time fields have to be functions of the gauge coordinates
$\xn$. This is forced on us when we require that it be possible to
define the gauge
transformation laws for the space time fields in a consistent way.
   This does not introduce any new physical degrees since these
can be gauge fixed. Nevertheless it is amusing to note that space-time
has effectively become infinite dimensional.

In order to make precise contact with string theory one has to perform another
step that we do not discuss in this paper. It involves generalizing to
the interacting case the dimensional reduction that was done in I (for
the free case).  Given that the basic technique involves calculation
of correlators of vertex operators on the world sheet we are more or
less guaranteed that we will reproduce bosonic string amplitudes. What
needs to be shown is that the dimensional reduction does not violate
the gauge invariance. We reserve this issue for a future publication
\cite{BSP}.

This paper is organized as follows.  In section II we give a short
review of \cite{BS2} and elaborate on the role of the parameter
$\s$. 
 In section
III we describe the generalization outlined in \cite{Puri}.  In section IV
we discuss the gauge invariance of the Loop Variable. 
Section V contains some examples of equations of motion. Section VI
discusses how one obtains the gauge transformation laws of the
space-time fields. Section VII discusses the consistency issue and
shows why the fields have to be functions of $\xn$. Section VIII
contains a summary and some concluding remarks. An Appendix contains
some details of a covariant Taylor expansion.

\section{Review}
\setcounter{equation}{00}
\subsection{Free theory}

In I the following expression was the starting point to obtain the
equations of motion at the free level: 
\be     \label{2.1}
e^A=e^{k_{0} ^{2}\Sigma + \sum _{n>0} k_{n} . k_{0} \dsn \Sigma + 
\sum _{n,m >0}k_{n} .k_{m}
(\dsnm  - \frac{\partial}{\partial x_{n+m}})\Sigma +ik_{n} Y_{n}}
\ee

The prescription was to vary w.r.t $\SI$ and evaluate at 
$\Sigma =0$ to get the 
equations of motion.  Here, 
$2 \Sigma \equiv < Y(z)Y(z) >$ and $Y = 
\sum _{n}\aln \frac{\partial^{n}X}{(n-1)!} \equiv
\sum _{n} \alpha _{n} \tilde{Y}_{n} $. 
$\aln $ are the modes
of the einbein $\al (t)$ used in defining the loop variable
\be \label {lv}
\lpp = \gvk
\ee
\[ \al (t) = \sum _{n \ge 0}\aln t^{-n}\]
\[ k(t) = \sum _{n\ge 0}\kn t^{-n}\]

 One can also show easily that 
$Y_{n} = \frac{\partial Y}{\partial x   _{n}}$.  
$\SI$ is thus a generalization of the Liouville mode, and what we
have is a generalization of the Weyl invariance condition on vertex 
operators.  

There is an alternative way to obtain the $\Sigma $ dependence \cite{BSCD}.
This is to perform a general conformal transformation on a vertex
operator by acting on it with $\eln $ using the relation 
\footnote{This relation is only true to lowest order in
$\lambda$.  The exact expression is given in 
\cite{BSV}}{\cite{BSV}:
\be  \label {2.2.5}
\eln e^{iK_{m} \tilde{Y}_{m}}
= e^{K_{n}.K_{m}\lambda _{-n-m}+\tilde{Y_{n}}\tilde{Y_{m}}\lambda_{+n+m}
+imK_{n}\tilde{Y_{m}}\lambda_{-n+m}}e^{iK_{m}\tilde{Y_{m}}}
\ee
The anomalous term is $\Kn .\Km \lambda _{-n-m}$ and the classical 
term is $ m\Kn \tilde{\ym} \lambda _{-n+m}$. We will ignore the 
classical piece: this can be rewritten as a $(mass)^{2}$ term, which 
will be reproduced by performing
a dimensional reduction, and other pieces involving derivatives
of $\Sigma$ (defined below) that correspond 
to field redefinitions \cite{BS1}.  
We can apply (\ref{2.2.5}) to the loop variable 
(\ref{lv}) by setting $\Km = \sum _{n} k_{m-n} \aln$.  Defining
\be
\Sigma =  \sum _{p,q} \alpha _{p} \alpha _{q} \lambda _{-p-q}
\ee
we recover (\ref{2.1}). It is the approach described above that 
generalizes more easily to the interacting case.

The equations thus obtained are invariant under 
\be   \label{2.3}
k_{n} \rightarrow \sum _{m}k_{n-m} \lambda _{m}
\ee
which is just the mode expansion of (\ref{1.1}). 

That this is an invariance of the equations of motion 
derived from (\ref{2.1}) follows essentially from the fact 
that the transformation (\ref{2.3}), applied to (\ref{2.1})
changes it by a total derivative.
\be  \label{2.4}
\delta A = \sum _{n} \lambda _{n} \dsn [A]
\ee

The equations are obtained by the operation
 ${\delta \over \delta \SI}A \mid _{\SI =0}$. 
Thus consider the gauge variation of this:
\[ \delta _{gauge}  {\delta \over \delta \SI}A =
 {\delta \over \delta \SI} \delta _{gauge}A\]
\[=  {\delta \over \delta \SI} \la _n \dsn A \] Now $A$ being linear
in $\SI$ and its derivatives can always be expressed after integration
by parts as $\SI B$ for some $B$. Thus we have 
\[ = {\delta \over \delta \SI}\la _n \dsn ( \SI B)= \la _n(-\dsn B
+\dsn B) =0\]

 Thus the equations obtained from (\ref{2.1}) are invariant.

The connection between these variables and transformation laws
and the usual fields and gauge transformations  was described in I. 
Briefly, the fields were defined by
\be     \label{2.41}
S_{n,m,...}^{\mu \nu ..}(\ko )=<\kn ^{\mu} \km ^{\nu}...>=\int [\prod _n
d\kn d\la _n]\kn ^{\mu} \km ^{\nu}...\Psi [\ko ,\ki ,\kt ,...,\kn,...\la _m ...]
\ee
where $\Psi$ is some ``string field'' that describes a given
configuration.

And the gauge parameters $\Lambda _{p,n,m..}^{\mu , \nu} (\ko )$ were
defined 
by a similar 
equation involving
one power of $\la _p$,  $p=1,2,...$, and arbitrary numbers of $\kn
,\km ..$ in the integrand. 
                      
However there are some caveats.  In proving (\ref{2.4})
one needs to use equations such as
\be \label{2.5}
\dsi (\dsii - \dst )\SI = (\dsic
-\dsit )\SI = 2(\dsit - \dsth ) \SI
\ee
which follow from the basic definitions \cite{BS1}. This implies that
equations of motion obtained by varying $\Sigma$ will not be invariant.
To see this consider the following expression:

\be     \label{2.33}
2(\dsit - \dsth )\SI A + ( \dsii - \dst )\SI 
\frac{\scriptstyle \partial A}
{\scriptstyle \partial x_{1}}
\ee
Using (\ref{2.5}) we get
\be     \label{2.34}
=\; \dsi [(\dsii - \dst )\SI A]
\ee
which is a total derivative.  However if we vary (\ref{2.33})
w.r.t. $\Sigma$, one gets
\be     \label{2.35}
2\delta \SI (\dsit + \dsth )A + 
\delta \SI (\dsii + \dst )  \frac{\scriptstyle \partial A}{
\scriptstyle \partial x_{1}}
\ee
which is not zero. On the other hand if we rewrite (\ref{2.33})
as (using (\ref{2.5}))
\be     \label{2.36}
(\dsic -\dsit )\SI A + (\dsii - \dst )\SI \frac{\scriptstyle \partial A}
{\scriptstyle \partial x_{1}}
\ee
and vary w.r.t $\Sigma$ we get
\be     \label{2.37}
\delta \SI (- \dsic - \dsit ) A + \delta \SI (\dsic + \dsit ) A
\ee
which is zero.

Thus one has to be careful about varying w.r.t $\Sigma$
indiscriminately. 
Let us review the solution to this as we will face the same issue in the
interacting case discussed in the next section.
Consider the variation of the exponent $A$ (\ref{2.1}), reproduced
below, due to $\la _p$: 
\be    
e^A=e^{k_{0} ^{2}\Sigma + \sum _{n>0}k_{n} . k_{0} \dsn \Sigma + 
\sum _{n,m>0}k_{n} .k_{m}
(\dsnm  - \frac{\partial}{\partial x_{n+m}})\Sigma +ik_{n} Y_{n}}
\ee
The change is 
\[
\la _p (\sum _n k_{n-p} . k_{0} \dsn \Sigma + \sum _{n,m\neq p}k_{n-p} .k_{m}
(\dsnm  - \dsq )\Sigma + 
\]
\be     \label{2.38}
+\sum _m k_m .k_0 (\frac{\partial ^2}{\partial x_m
\partial x_p}- \frac{\p}{\p x_{m+p}})\Sigma
+i\sum _n k_{n-p} Y_{n})
\ee
If we assume tracelessness of the gauge parameter so that any term of
the form $\la _p \kn .\km $ is zero then the second sum in 
(\ref{2.38}) vanishes and using the fact that the first sum cancels
the second term in the last sum we can rewrite the variation of $A$ as 
\[
\la _p \frac{\partial}{\partial x_p} \{ \sum _m \km . \ko
 \frac{\partial}{\partial x_m}\SI + \sum _n ik_{n-p}Y_{n-p}+ \sum
 _{n,m} \kn . \km (\dsnm -\dsq )\SI\}  
\]
\be     \label{2.39}
=\la _p \frac{\p}{\p x_p}A
\ee
Note that in the first line of this equation we have added a term that
vanishes by the tracelessness constraint, {\em viz}
 terms 
involving $\la _p \kn . \km$ . But it is important that we
have {\em not} used identities of the type given in (\ref{2.5}). Thus
tracelessness of the gauge parameters ensures the gauge invariance of
the equations.

\subsection{Interactions}   

In II this approach was generalized to include some interactions.  
The basic idea was to introduce a new parameter 
$\sigma : 0\leq \sigma \leq 1$
to label different strings and to replace each $\kn$ in the free 
equation by $\int _{0}^{1} d \sigma \kn (\sigma )$. 
The next step was to assume
that 
\be  \label{2.7}
<\kim (\sigma _{1})\kin (\sigma _{2})> = S^{\mu\nu}\delta 
(\sigma _{1}-\sigma_{2})
+ A^{\mu}A^{\nu}
\ee
where $< ...>$ denotes $\int {\cal D}k(\sigma ) ...\Psi [k(\sigma )]$, 
$\Psi $ being
the``string field''  defined in I.\footnote{No special property of 
$\Psi $ is assumed other than this.} This corresponds to saying that when 
$\sigma _{1} = \sigma _{2}$, both the $\ki $'s belong to the same string
and otherwise to different strings where they represent two 
photons at an interaction point.\footnote{It will be seen that   
(\ref{2.7})has to be generalized by replacing the $\delta$-function
on the RHS by something else, when we attempt to reproduce string
amplitudes \cite{BSPT}.
However in this paper we will not do so.}
The gauge transformation is replaced 
by (\ref{1.2}).  This is easily seen to give interacting interacting
equations.  However the fact is that this is only a leading term
in the infinite set of interaction vertices.

As a prelude to generalizing this construction, let us explain
more precisely the nature of the replacement $\kn \rightarrow
\int _{0}^{1} d \sigma \kn ( \sigma )$.  Let us split the interval
$(0,1)$ into $N$ bits of width $a=\frac{1}{N}$.  We will
assume that when $\s$ satisfies $\frac{n}{N} 
\leq \sigma \leq \frac{n+1}{N}$
it represents the $(n+1)$th string.  Let us also define a function
\br \label{2.8} 
D(\sigma _{1},\sigma _{2})& =&1 \;  if\; 
\sigma _{1},\sigma_{2}\; belong\; to\; the\;
same\; interval     \nonumber    \\
&=&0 \;if\;\sigma_{1}\;,\sigma_{2} \;belong\;to\;different\;intervals.
\er
Thus $\int _{0}^{1} d \sigma _{1}D(\sigma _{1} \sigma _{2})  = \; a \; =
\int _{0}^{1} d \sigma _{1}\int _{0}^{1} d 
\sigma _{2}D(\sigma _{1} \sigma _{2})$.

Then we set 
\be   \label{2.9}
<k^{\mu}(\sigma_{1})k^{\nu}(\sigma_{2})> = 
\frac{D(\sigma_{1},\sigma_{2})}{a}S^{\mu \nu}   
+A^{\mu}A^{\nu}
\ee
In the limit $N \rightarrow \infty ,\; a \rightarrow 0, \; 
\frac{D(\sigma _{1} , \sigma _{2})}{a}  
\approx \delta (\sigma _{1} -\sigma _{2})$
and we recover (\ref{2.7}). 

In effect (\ref{2.1}) has been modified to
\be  \label{2.10}
e^{\int_{0}^{1}\int_{0}^{1}  d\sigma_{1}d\sigma_{2}
[k_{0}(\sigma_{1})k_{0}(\sigma_{2})\Sigma
+k_{n}(\sigma_{1}).k_{0}(\sigma_{2})\dsn \Sigma + \sum _{n,m}
(\dsnm - \frac{\partial}{\partial x_{n+m}})\Sigma ] +\int _{0}^{1}
d\sigma k_{n}(\sigma )Y_{n}}
\ee
The final step (which is also necessary in the free case), is to
dimensionally reduce to obtain the massive equations.  For details
we refer the reader to I.

The modification (\ref{2.7}), that replaces $S^{\mu \nu}$ by
$S^{\mu \nu} + A^{\mu} A^{\nu}$ can be understood in terms of
the OPE.  Consider a correlation function involving two vector
vertex operators and any other set of operators, that we represent as
\be  \label{2.11}
{\cal A}=<V_{1}V_{2}...V_{N}:\kim \partial_{z}X^{\mu}\e 
:\qin \partial_{w}X^{\nu}
 \qe>
\ee
The OPE of 
$:\kim \partial_{z}X^{\mu}(z) \e :$ and $: \qin \partial_{w} 
X^{\nu}(w)\qe :$

is given by 
\[
:\kim \partial_{z}X^{\mu}(z) \e :: \qin \partial_{w} X^{\nu}(w)\qe : =
\]
\be  \label{2.12}      
:\kim \qin \partial_{z} X^{\mu} \partial_{w} X^{\nu} 
e^{i(k_{0}X(z) + q_{0}X(w))}:
+\; terms\; involving\; contractions.
\ee

We can Taylor expand
\be     \label{2.13}
X(w) = X(z) + (w-z)\partial _{z} X + O(w-z) +...
\ee

This gives for the leading term in (\ref{2.11})
\be     \label{2.14}
{\cal A} = <V_{1}V_{2}...V_{N}:\kim \qin \partial_{z} X^{\mu} 
\partial_{z} X^{\nu} e^{i(k_{0}X(z) + q_{0}X(w))}:>
\ee
Compare this with the correlation involving $S^{\mu \nu}$:
\be     \label{2.15}
{\cal A'} = <V_{1}V_{2}...V_{N}:\kim \kin \partial_{z} X^{\mu} 
\partial_{z} X^{\nu} \e:>
\ee
We see that ${\cal A} $ and ${\cal A'}$ give identical terms except
that $S^{\mu \nu}$ is replaced by $A^{\mu}A^{\nu}$.  It is 
in this sense that the substitution given in II, gives the leading term 
in the OPE.  The crucial point is that, while in (\ref{2.10}) we have
introduced the parameter $\sigma $ in the $k _{n}$'s we have not 
done so for the $\yn$'s.  This is equivalent to approximating
$X(w)$ by $X(z)$ in (\ref{2.13}).  Clearly, the generalization required
to get all the terms is to introduce the parameter $\s$ in $Y$ also.
We turn to this in the next section.
\section{Interactions}
\setcounter{equation}{00}
\subsection{Introducing $\s$-dependence in the loop variable}

We will introduce the parameter $\s$ in all the variables keeping
in mind the basic motivation that $\s$ labels different vertex 
operators.  Thus all the variables that are required to define
a vertex operator become $\s$ dependent.
Thus
\be     \label{3.1}
X^{\mu}(z)\rightarrow X^{\mu}(z(\sigma ))
\ee
\be     \label{3.2}
\xn \rightarrow \xn (\sigma )
\ee
in addition to
\be     \label{3.3}
\kn ^{\mu}\rightarrow \kn ^{\mu}(\sigma )
\ee
The $\sigma$-dependence of $\xn$ in eqn. \ref{3.2} is only an
intermediate step. At the end of the day (but before any integration
by parts is done) we will set all the $\xn$'s to be the same. One can
think of this merely as a device for keeping track of which term is
being differentiated.

(\ref{3.1}) and (\ref{3.2}) imply that
\be     \label{3.4}
\frac{\partial}   {\partial\xn} Y \rightarrow 
\frac{\partial}{\partial \xns}
Y(z(\sigma ),\xns ) 
\ee
Note that $X$ need not be an explicit function of $\s$ since at a given
location $z$, on the world sheet there can only be one $X(z)$.  As an
example of the above consider the case when we have regions $(0,1/2)$
and $(1/2,1)$.  When $0\leq \sigma \leq 1/2$ one has $z(\sigma ) \equiv z$
and for $1/2 \leq \sigma \leq 1$ one has $z(\sigma ) \equiv w$.  Similarly
$x_{n}(\sigma )$ could be called $x_{n},y_{n}$ in the two regions
and $\kn (\sigma )$ could be called $\kn , p_{n} $ in the two regions.
Thus in this example the vertex operator 
$\kn (\sigma )\yn (z(\sigma ), x_{n}( \sigma ))
e^{ik_{0}(\sigma )Y(\sigma )}$
stands for $\kn \frac{\partial Y}
{\partial x   _{n}}(z,x_{i})e^{ik_{0}Y(z,x_{n})}$
and $p_{n} \frac{\partial Y}{\partial y_{n}}(w,y_{i})
e^{ip_{0}Y(w,y_{n})} $ in the
two regions.

 Now we have to clarify what we
mean by a derivative w.r.t $x_{n}(\sigma )$:  In (\ref{3.4}) we have 
$\frac{\partial Y(z(\sigma ), x_{i}(\sigma ))}{\partial\xns}$ : 
One has to specify
the meaning of $\frac{\partial\xns }{\partial\xnsp}$.  
Clearly what we want is:
If $\sigma , \sigma '$ belong to the same interval, then 
$\frac{\partial\xns }{\partial\xnsp} \; = \; 1$ and zero otherwise. 
Thus using
(\ref{2.8})
\be     \label{3.5}
\frac{\partial\xns}{\partial\xnsp} = D(\sigma , \sigma ')
\ee
or more generally
\be     \label{3.6}
\frac{\partial\xns}{\partial\xmsp} = \delta _{nm} D(\sigma ,\sigma ')
\ee
Note that this is not the same as the conventional functional
derivative.  However we can define
\be     \label{3.7}
\frac{\delta \xns }{\delta \xnsp } \equiv \frac{D(\sigma ,\sigma ')}{a}
\ee
which, in the limit $a\rightarrow 0$ becomes the usual
functional derivative.  Thus
\be     \label{3.8}
\int d \sigma ' \frac{\delta Y (\sigma )}{\delta \xnsp} 
= \frac{\partial Y(\sigma )}{\p
\xns}
\ee
We can now write down the generalization of (\ref{2.1})
\[
exp \{\int \int d\sigma_{1} d \sigma _{2} 
\{ k_{0}(\sigma _{1}).k_{0}(\sigma _{2})
[\tS (\sigma _{1},\sigma _{2}) + \tilde{G}(\sigma_{1},\sigma _{2})]
\]
\[
+\int \int d\sigma _{3} d\sigma _{4}\sum _{n,m \ge 0} 
\kn (\sigma _{1}).\km (\sigma _{2})
\]
\[
\frac{1}{2}[\frac{\delta ^{2}}{\delta \xn (\sigma _{1})
\delta \xm (\sigma _{2})}
-\delta (\sigma _{1}-\sigma _{2})\frac{\delta }
{\delta x_{n+m}(\sigma _{1})}]
[\tS(\sigma _{3},\sigma _{4}) + \tilde{G}(\sigma _{3},\sigma _{4})]\}\}
\]
\be     \label{3.9}  
exp \{i\int d \sigma k_{n}(\sigma )Y_{n}(\sigma )\}
\ee

For convenience of notation have assumed the following:
\[{\delta \aln (\si )\over \delta x_0(\st)}={D (\si - \st )\over a}\aln
(\si )\]
This saves us the trouble of writing separately the case $n=0$
in the sum in (\ref{3.9}). 

In (\ref{3.9}) $G(\sigma _{1} ,\sigma _{2})=
\tilde{G}(z(\sigma _{1}), z(\sigma _{2}))$
=$<Y(z(\sigma _{1}))Y(z(\sigma_{2}))>$ is the Green function which starts
out as $ln (z_{1} - z_{2})$.  We have suppressed the Lorentz indices.
One might expect by Lorentz invariance
 $\tilde{G}^{\mu \nu} = \delta ^{\mu \nu} \tilde{G}$.  However in I it
was seen that the $D+1$th coordinate has a special role and is like the
ghost coordinate of bosonic string theory. So there is no reason to
expect the full $SO(D+1)$ invariance. In fact \cite{BSP} we will
have to assume some specific properties for $\tG ^{D+1,D+1}$ in order
to reproduce string amplitudes.

More precisely, if we define:
[Using the notation $z_{i} = z(\sigma _{i})$]
\be     \label{3.10}
D_{z_{1}} = D_{z(\sigma _{1})} \equiv 1+\ai 
(\sigma _{1})\frac{\partial}
{\partial z (\sigma _{1})} + \at \frac{\pp}
{\partial z^{2} (\sigma _{1})}+...
\ee
so that
\be    \label{3.11}
Y(z(\sigma ))=D_{z(\sigma )}X(z(\sigma ))
\ee
then,
\be     \label{3.12}
\tilde{G}(z_{1},z_{2}) = D_{z_{1}} D_{z_{2}}G(z_{1},z_{2})
\ee
\be     \label{3.13}
\tS(\sigma _{1},\sigma_{2}) =  D_{z_{1}} 
D_{z_{2}}\rho (\sigma _{1},\sigma _{2})
\ee
where 
\be     \label{3.14}
\rho (\sigma _{1} , \sigma _{2})=
\frac{\la (z(\sigma _{1}))-\la (z(\sigma _{2}))}
{z(\sigma _{1})-z(\sigma _{2})}
\ee
is the generalization of the usual Liouville mode $\rho (\sigma )$
which is equal to $\frac {d \lambda }{d z}$.  The $\tilde {\Sigma}$
dependence in (\ref{3.9}) is obtained by the following step:
\be     \label{3.15}
e^{:\frac{1}{2}\int du \lambda (u) [\partial _{z} X(z+u)]^{2}:}
e^{ik_{n}\frac{\partial}{\partial x_{n}} D_{z_{1}}X}
e^{ip_{m}\frac{\partial}{\partial x_{m}}D_{z_{2}}X}
\ee
defines the action of the Virasoro generators on the two sets of 
vertex operators.
\be     \label{3.16}
= e^{ik_{n}.p_{m}\partial _{x_{n}}\partial _{y_{m}}D_{z_{1}}D_{z_{2}}
\oint du\frac{\lambda (u)}{z_{1}-z_{2}}[\frac{1}{z_{1}-u} -\frac{1}{z_{2}-u}]}
\ee
\be     \label{3.18}
=e^{ik_{n}.p_{m}\partial _{x_{n}}\partial _{y_{m}} \tilde{\Sigma}}
\ee     
This expression is only valid to lowest order in $\lambda $
which is all we need here.\footnote{The exact expression is given in
\cite{BSV}}.
The expression 
\be     \label{3.20}
\int \int d\sigma _{1} d\sigma _{2}
\frac{1}{2}[\frac{\delta ^{2}}
{\delta \xn (\sigma _{1})\delta \xm (\sigma _{2})}
-\delta (\sigma _{1}-\sigma _{2})
\frac{\delta }{\delta x_{n+m}(\sigma _{1})}]
[\tS(\sigma _{3},\sigma _{4}) + \tilde{G}(\sigma _{3},\sigma _{4})]
\ee
can easily be seen to be equal to
\be     \label{3.21}
\frac{\pp}{\partial\xn (\sigma _{3}) 
\partial\xm (\sigma _{3})}\tS (\sigma _{3},\sigma _{4})         
\ee
In the limit $\sigma _{3} = \sigma _{4}=
\sigma $ this is just equal to
$1/2[\frac{\pp}{\partial\xns \partial\xms} - 
\frac{\partial}   {\partial x_{m+n}(\sigma )}]
\tilde{\Sigma} (\sigma , \sigma )$ and 
reduces to the free field case described
by (\ref{2.10})(provided the limit is taken after differentiation).

Let us show  that the gauge transformation (\ref{1.2})
changes (\ref{3.9}) by a total derivative
\be     \label{3.22}
\delta A = \int d\sigma ' \la ( \sigma ' )  \int d\sigma
\frac{\delta }{\delta \xn (\sigma )}A
\ee

\section{Invariance of the Loop Variable}
\setcounter{equation}{00}

Our starting point is the loop variable $e^A$ given by:

\eA

Under a gauge transformation:

\be 
\kn (\si ) \rightarrow \int d\s \la _p (\s ) k_{n-p} (\si )
\ee 

Let us consider $p=1$.
\be
\ki (\si ).\ko (\st ) \dsis [\tS +\tG ](\si ,\st ) \rightarrow \int d\s
\li (\s )  \ko (\si ) . \ko (\st ) \dsis [\tS +\tG ](\si , \st )
\ee
\be
\ko (\si ).\ki (\st ) \dsis [\tS +\tG ](\si ,\st ) \rightarrow \int d\s
\li (\s )  \ko (\si ) . \ko (\st ) \dsts [\tS +\tG ](\si , \st )
\ee

Adding the two we get:
\be
\int d\s
\li (\s )[\dsis + \dsts ]  \ko (\si ) . \ko (\st ) [\tS +\tG ](\si ,
\st ) 
\ee

Similarly,
\be
\kt (\si ).\ko (\st ) \dsix [\tS +\tG ](\si ,\st ) \rightarrow \int d\s
\li (\s )  \ki (\si ) . \ko (\st ) \dsix [\tS +\tG ](\si , \st )
\ee

\be
\ko (\si ).\kt (\st ) \dstx [\tS +\tG ](\si ,\st ) \rightarrow \int d\s
\li (\s )  \ko (\si ) . \ki (\st ) \dstx [\tS +\tG ](\si , \st )
\ee

\be
\ki (\si ).\ki (\st ) \dsb  \rightarrow \int d\s
\li (\s )  (\ki (\si ) . \ko (\st ) + \ko (\si ).\ki (\st ))\dsb
\ee

Adding we get,

\[
\int d\s \li (\s ) \{[\dsis + \dsts ]\ki (\si ) .\ko (\st ) \dsis [\tS
+ \tG ]+ \]
\be [\dsis + \dsts ]\ko (\si ) .\ki (\st ) \dsts [\tS + \tG ]\}
\ee

\be
=\int d\s \li (\s ) [\dsis + \dsts ]\{\ki (\si ) .\ko (\st ) \dsis [\tS
+ \tG ] + \si \leftrightarrow \st \}
\ee

Now consider

\be
\kt (\si).\ki (\st )\frac{\pp [\tS +\tG ]}{\p x_2(\si )\p x_1(\st )}+ 
\ki (\si).\kt (\st )\frac{\pp [\tS +\tG ]}{\p x_1(\si )\p x_2(\st )}
\ee

\be 
\rightarrow
\int d\s \li (\s ) \ki (\si ).\ki (\st )
\frac{\pp [\tS +\tG ]}{\p x_2(\si )\p x_1(\st )}+
\int d\s \li (\s ) \ki (\si ).\ki (\st )
\frac{\pp [\tS +\tG ]}{\p x_1(\si )\p x_2(\st )}
\ee 

\be 
= \int d\s \li (\s ) [\dsis +\dsts ]\ki (\si ).\ki (\st )
\dsb
\ee

From the above it is clear that we get the following:
\be
\delta A = \int d\s \li (\s )[\dsis + \dsts ] A 
\ee
On setting $\xn (\si )= \xn (\st ) = \xn $ we get
\be
\delta A = \li (\s ) \dsi A 
\ee

Thus to lowest order in $\li $, $e^A$ changes by a total derivative in
$x_1$. This is obviously true for $\la _p$ also. 

Thus the equations obtained by varying w.r.t $\tS (z(\si ),z(\st ),
\xn (\si ),\xn (\st ))$ are
invariant. However
$\tS$ is not a local field on the world sheet. $A$ has terms of the form
$[{\p ^2 \over \p x \p y }\SI (w,z,x,y)]\mid _{x=y} \ne {\p ^2 \over \p x^2}[
\SI (w,z,x,y) \mid _{x=y}]$. Thus $A$ cannot be expressed in terms of 
$\xn $-derivatives of a field.  We would have to use both $\xn$ and $\yn$.
But we cannot integrate by parts on both $\xn ,y_n$ - there is no
such gauge invariance.
 So we first
Taylor expand it in powers of $z(\st ) - z(\si )$ the coefficients of
which are derivatives of a local field $\SI (z,x )\equiv \bar \SI (z,x,y)  \mid
_{x=y}$, where $\bar \SI (v,x,y) = \tS (v,v,x, y)$. Below we have used
the letter $v$ to denote $z(\si ) + z(\st ) \over 2$ and $x(\si )=x ,
x(\st )=y$.

\be
\tS (\si , \st ) = \bar \SI (v) + a D_1(x,y)\bar \SI (v) + a^2 D_2
(x,y)\bar \SI (v) + ...
\ee

\be     
=
\bar{\Sigma} (v) + a
\sum_r(\gamma_r'^0 \frac{\partial\bar{\SI}}{\partial y_{r+1}}
- \gamma _r^0\frac{\partial\bar{\SI}}{\partial x_{r+1}})
+
\ee
\[
 {a^2\over 2!}[\sum _s (\gamma_s'^1 \frac{\partial\bar{\SI}}{\partial
y_{s+1}}+ \gamma _s ^1\frac{\partial\bar{\SI}}{\partial x_{s+1}})
-2\sum _{r,s}\gamma _r ^0 \gamma _s '^0\frac{\partial
^2\bar{\SI}}{\partial x_{r+1}
\partial y_{s+1}}] + ...
\]

$D_k$ and $\gamma $ are defined in the Appendix. Very explicitly, the
first few terms of the Taylor expansion are :
\[
\tS (\si , \st ) = \bar \SI (v) + a\underbrace {
[\frac{\p \bar \SI }{\p y_1}-\frac{\p
\bar \SI}{\p x_1} + ({y_1^2\over 2}+ y_2){\p \bar \SI \over\p y_3}-({x_1^2
\over 2} + x_2 ) \frac{\p\bar \SI}{\p x_3} ]}_{D_1(x,y)} +
\]
\be
{a^2 \over 2}\underbrace  { 
[
\frac{\p \bar \SI}{\p x_2} +\frac{\p \bar \SI}{\p y_2}
-2\frac{\pp \bar \SI}{\p x_1 \p y _1} + x_1 \frac{\p \bar \SI}{\p
x_3}+
y_1\frac{\p \bar \SI}{\p y_3} -
2({x_1^2\over 2}+ x_2){\pp \bar \SI \over \p x_3 \p y_1}-
2({y_1^2\over 2}+ y_2){\pp \bar \SI \over \p y_3 \p x_1}
]}_{D_2(x,y)}+...
\ee

Once you Taylor expand $\SI$ we have the following problem that we
encounter also in the free case. The problem is that when the gauge
variation does not produce $\ko$ we need constraints:
\[
\kt .\ki [ \frac{\pp \SI}{\p x_2 \p x_1}- \frac{\p \SI}{\p x_3}]
\rightarrow \li \ki .\ki [ \frac{\pp \SI}{\p x_2\p x_1}- \frac{\p
\SI}{\p x_3}] \] 
\be ?=? \li \ki .\ki \dsi [\frac{\pp \SI}{\p x_1^2}-\frac{\p \SI}{\p
x_2}]
\ee
The last equation is not an identity and follows only because certain
constraints are obeyed by $\SI$. This in turn  requires the imposition of the
constraints on the gauge parameters - $\li \ki .\ki =0$.

This problem is not there for the
$\lt$ variation as can be seen in the following:

\[
\kt .\ki [ \frac{\pp \SI}{\p x_2 \p x_1}- \frac{\p \SI}{\p x_3}]
\rightarrow \lt \ko .\ki
 [ \frac{\pp \SI}{\p x_2 \p x_1}- \frac{\p \SI}{\p x_3}]
\]

\[
{k_3.\ko \over 2}\frac{\p \SI}{\p x_3} \rightarrow {\lt \ki .\ko \over
2} \frac{\p\SI}{\p x_3}\]

They add up to:
\[ \lt \kt .\ko {\pp \SI \over \p x_2 \p x_1}=\lt \dst [{\ki .\ko
\over 2}{\p \SI \over \p x_1}]\] 

For the above argument to go through in the interacting case we need
the following property for the Taylor expansion coefficients:
\be
\dsnm [D_k (x,y) \bar \SI ] = \dsq [D_k (x,y)\bar \SI ]
\ee
(and the same obviously for $\yn$). It is demonstrated in the Appendix
that this is in fact true.
  
Thus in general consider:
\be
\kn (\si ). \km (\st ) 
[{\pp D_k(x,y)\bar \SI \over \p \xn \p y_m}]\mid _{x=y}
+k_{n+m}(\st ).\ko (\si )
[{\p D_k(x,y)\bar \SI \over \p y_{n+m}}]\mid _{x=y}
\ee

\be
\rightarrow
\int d \s \la _n (\s ) \ko (\si ).\km (\st ) 
[{\pp D_k(x,y)\bar \SI \over \p \xn \p y_m}]\mid _{x=y}+
\int d\s \la _n(\s ) \km (\st ).\ko (\si )
[{\p D_k(x,y)\bar \SI \over \p y_{n+m}}]\mid _{x=y}
\ee
\be
=\int d\s \la _n(\s )\ko (\si ).\km (\st )[ (\dsn +{\p \over \p y_n})
{\p D_k(x,y)\bar \SI \over \p y_{m}}]\mid _{x=y}
\ee
\be
=\int d(\s ) \la _n (\s ) \dsn [\ko (\si ). \km (\st )
{\p D_k(x,y)\bar \SI \over \p y_{m}}\mid _{x=y}]
\ee

Similarly,
\be
\kn (\st ). \km (\si ) 
[{\pp D_k(x,y)\bar \SI \over \p \xm \p y_n}]\mid _{x=y}
+k_{n+m}(\si ).\ko (\st )
[{\p D_k(x,y)\bar \SI \over \p x_{n+m}}]\mid _{x=y}
\ee

\be
\rightarrow
\int d \s \la _n (\s ) \ko (\st ).\km (\si ) 
[{\pp D_k(x,y)\bar \SI \over \p \xm \p y_n}]\mid _{x=y}+
\int d\s \la _n(\s ) \km (\si ).\ko (\st )
[{\p D_k(x,y)\bar \SI \over \p x_{n+m}}]\mid _{x=y}
\ee
\be
=\int d\s \la _n(\s )\ko (\st ).\km (\si )[ (\dsn +{\p \over \p y_n})
{\p D_k(x,y)\bar \SI \over \p x_{m}}]\mid _{x=y}
\ee
\be
=\int d(\s ) \la _n (\s ) \dsn [\ko (\st ). \km (\si )
{\p D_k(x,y)\bar \SI \over \p x_{m}}\mid _{x=y}]
\ee

Adding the two we find that the $\la _n$ variation is a total
derivative in $\xn$ of $A$ even after Taylor expanding.

Similarly the tracelessness constraint of the free theory generalizes
to
\be
< \int d(\s ) \la _p (\s )[\kn (\si ).\km (\st ) ] ....>=0
\ee
in the equation of motion. All the above guarantees that the variation
of $e^A$ is a total derivative and therefore the equations of motion
obtained by varying w.r.t $\SI$ are invariant.

\section{Examples}
\setcounter{equation}{0}

\subsection{Vector  $\ki $ Contribution to $\yim$}
 Our starting point is $e^A$ given by \eA 
We keep terms with one $\ki$ only. There are three terms that contribute.

\be
e^{\int \ko (\s _5 ).\ko (\s _6)[\tS + \tG ]}
\ki (\si ).\ko (\st ) \dsis [\tS +\tG ] \e +
e^{\int \ko (\s _5 ).\ko (\s _6)[\tS + \tG ]}
i\int \ki Y_1 \e
\ee 

In leading order we have:
\[
 \dsis \tS (\si ,\st ) = 
{\p \bar \SI \over \p x_1} = {1\over 2}{\p \SI \over \p x_1}
\]
\be     \label{Approx}  
 \tS = \bar \SI = \SI \ee
We get 
\be
\ki (\si ).\ko (\st ) {\p \bar \SI \over \p x_1}\e +
\ki (\st ).\ko (\si ) {\p \bar \SI \over \p y_1}\e +
\ko (\si ).\ko (\st )\bar \SI i\int \ki Y_1 \e
\ee
which on setting $\xn =y _n$ becomes
\be
=
{\ki (\si ).\ko (\st )  + \ki (\st ).\ko (\si )]\over 2} 
{\p \SI \over \p x_1}\e + \ko (\si ).\ko (\st ) \SI i\int \ki Y_1
\ee
Setting $\delta \over \delta \SI$ of this expression to zero we get
the equation (we can set all the $\sigma$ 's to be equal)
\be     \label{15}
-\ki (\si )\ko(\si ) i\kom \yim + \ko(\si ).\ko(\si )i\kim \yim =0
\ee
Converting to space-time fields the coefficient of $\yim$ is:
\be
= -\mup \p ^{\nu} A^{\mu} + \mup \p ^{\mu} A^{\nu} = \mup F^{\mu \nu}=0
\ee
which is Maxwell's equation. (\ref{15}) is clearly invariant under 
$\ki (\si ) \rightarrow \ki (\si )+ \int d \s \la _1 (\s )\ko (\si )$, which in
 terms of space-time fields is $A_\mu \rightarrow A_\mu + \p _\mu \Lambda$.

\subsection{$\ki \ki$ and $\kt$ Contribution to $\yim$}
(i)
\[
{1\over 2!}
\{\ki(\si ).\ko(\st )\dsis [\tS +\tG ] +\si \leftrightarrow \st \}
\]
\be
\{\ki(\s _3 ).\ko(\s _4 ){\p \over \p x_1(\s _3 )} [\tS +\tG ] +\s _3 \leftrightarrow \s _4 \}
e^{\int \ko (\s _5 ).\ko (\s _6)[\tS + \tG ]}\e
\ee
(ii)
\be
\ki (\si ).\ki(\st )
\dsb e^{\int \ko (\s _5 ).\ko (\s _6)[\tS + \tG ]}\e
\ee
(iii)
\be
e^{\int \ko (\s _5 ).\ko (\s _6)[\tS + \tG ]}
\{\ki(\si ).\ko(\st )\dsis [\tS +\tG ] +\si \leftrightarrow \st
\}i\int \ki Y_1 \e
\ee

Let us consider each in turn:

\noindent
{\bf (i)}
 
Using the leading order expressions given in (\ref{Approx}) we get
\be
2 \times {1\over 2!}
e^{\int \ko (\s _5 ).\ko (\s _6)[ \tG ]}
\ki (\si ).\ko(\st ){\p \SI \over \p x_1}
[2
 \ki (\s _3).\ko (\s _4) {\p \tG (\s _3,\s
_4)\over \p x_1 (\s _3)}]\e
\ee

Varying w.r.t $\SI$ gives
\be
= -2
e^{\int \ko (\s _5 ).\ko (\s _6)[ \tG ]}
\ki (\si ).\ko(\st ) \ki (\s _3).\ko (\s _4) {\p \tG (\s _3,\s
_4)\over \p x_1 (\s _3)} i\int \ko {\p Y\over \p x_1}\e
\ee

Now we have to consider all possible contractions of the $\kn$'s. In
order to keep track of the possibilities we separate them into two
cases: those involving only one point on the world sheet (i.e. only
one vertex operator) and those involving two distinct points (two vertex
operators). 

\noindent
{\bf One Vertex Operator}

In the first case we have to be careful about
regularizing. Let us refer to the point as $\s _A$ with $z(\s _A )=z$ as
the location of the vertex operator. When there is a need for
regularizing we will let $\s_B$ be the second point with $z(\s
_B)-z(\s _A) = \epsilon$. We now let the various $\s _i$ be equal to 
$\s _A$ or $\s _B$ in all possible combinations, but we divide by 2
since these are actually the same point. 

Following this procedure we
see that regularization is required for $\s _3 , \s _4$ when they
stand for the same point (and also for
$\s _5 ,\s _6$, which we ignore for the moment). Thus we can let \[\s _3
= \s _A , \; \s _4 = \s _B\] and $\si ,\st$ can be anything.
This gives 
\be
-2 \ki (\s _A).\ko (\s _A) \ki (\s _A).\ko (\s _B){1\over z_A - z_B}
\ee
Now we let $z_A \rightarrow z_B$ and $\s _A \rightarrow \s _B$ and use:
\be     \label{S11}
<\kim (\s _A)\kin (\s _A)> = 
<\kim (\s _A)\kin (\s _B)> = 
S^{\mu \nu}_{1,1}(\ko )
\ee
This gives
\be     \label{24}
2S_{1,1}^{\mu \nu }\kom \kon {1\over \epsilon}.
\ee
 The other possibility is
 \[ \s _3 = \s _B , \; \s _4 = \s _A\] and again $\si ,\st$ can be
anything, 
which gives
\be
-2 \ki (\s _A).\ko (\s _A) \ki (\s _B).\ko (\s _A){1\over z_B - z_A}
\ee
Using (\ref{S11}) gives
\be     \label{26}
2S_{1,1}^{\mu \nu }\kom \kon {-1\over \epsilon}
\ee

Adding the two ((\ref{24}) and (\ref{26})) we get zero.

\noindent
{\bf Two Vertex Operators}

Now we go to the second possibility viz. there are two distinct
points. Let us call them $\s _I$ and $\s _{II}$ and let $z(\s _I)=z$
and $z(\s _{II})=w$. Thus we can have a)$\si = \s _I$ and $\s _3 = \s
_{II}$ or b) vice versa.

Consider a): 

First we consider the case that does not require regularization.

\noindent
{\bf Non-singular Case}
 
\be
-2 \ki (\s _I).\ko (\st) \ki (\s _{II}).\ko (\s _4){1\over w - z (\s _4)}
\ee
Let $\s _4 = \s _I$ and $\st = \s _I \; or \; \s _{II}$ 

We now use the notation $\ko (\s _I)=p$ and $\ko (\s _{II}) = q$. Thus
\[
<\kim (\s _I)> = A ^{\mu}(p)\]
\be     \label{A1}
<\kim (\s _{II})>= A^{\mu}(q)
\ee and
we get as contribution to the equation of motion:
\be     \label{Aa}
\int dz \int dw A^{\mu}(p)(p+q)^{\mu}A^{\nu}(q)p^{\nu}{1\over w-z}
\ee
We have explicitly written out the integrals over $z$ and $w$ to
emphasize the symmetry. Thus
by antisymmetry of the integrand in $z,w$ this is zero.

\noindent
{\bf Singular Case}

If we let $\s _4 = \s _{II}$ (and $\st = \s _I \; or \; \s _{II}$) we have to
regularize. 
So we split $\s
_{II}$ into $\s _A$ and $\s _B$ as before. Again either 
\[\s _3 = \s _A\; and \; \s_4= \s_B\] which gives 
\be
-2 \ki (\s _I).\ko (\st) \ki (\s _A).\ko (\s _{B}){1\over w_A - w_B}
\ee
and using (\ref{A1})
\be
=A^{\mu}(p)(p+q)^{\mu}A^{\nu}(q)q^{\nu}({-1\over \epsilon})
\ee 

or

\[\s _3 = \s _B\; and \; \s_4= \s_A\] which gives 
\be
-2 \ki (\s _I).\ko (\st) \ki (\s _B).\ko (\s _{A}){1\over w_B - w_A}
\ee
and using (\ref{A1})
\be
=A^{\mu}(p)(p+q)^{\mu}A^{\nu}(q)q^{\nu}({+1\over \epsilon})
\ee 

Adding the two contributions again gives zero.

We have also to look at possibility b) which was $\si = \s _{II}$ and
$\s _3= \s _{II}$
Analogous to (\ref{Aa}) one gets
\be     \label{Ab}
\int dz \int dw A^{\mu}(q)(p+q)^{\mu}A^{\nu}(p)p^{\nu}{1\over z-w}
\ee

Note that this is (upto a sign) 
(\ref{Aa}) with the labels $p,q$ interchanged. But $p,q$
being integration variables we get back (\ref{Aa}) but the overall
sign being opposite, 
 they cancel. The integration $\int \int dzdw$ also ensures the
vanishing of each term , viz (\ref{Aa}) and (\ref{Ab}), individually. 
This is also as it should be since
interchanging $z$ with $w$ is equivalent to interchanging momenta.

{\bf (ii)}
\be
\ki (\si ).\ki(\st )
{\pp \tS (\si ,\st )\over \p x_1(\si )\p x_1(\st )}
 e^{\int \ko (\s _5 ).\ko (\s _6)[ \tG ]}\e
\ee

Use
\[
{\pp \tS (\si ,\st )\over \p x_1(\si )\p x_1(\st )}=
{\pp \bar \SI \over \p x_1 \p y_1} +...
\]
On setting $\xn = y_n$,
\be
= {1\over 2}({\pp \SI \over \p x_1^2} - {\p \SI \over \p x_2}) +...
\ee

Only the first term can give, on integration by parts, a contribution
to $Y_1$:
\be
 e^{\int \ko (\s _5 ).\ko (\s _6)[ \tG ]}
 \int \ko (\s _7 ).\ko (\s _8) {\p \tG (\s _7 , \s _8)\over \p x_1}  
\ki (\si ).\ki(\st )i\ko Y_1 \e
\ee
\be
\tG (\s _7, \s _8 )= ln |z(\s _7)-z(\s _8)| +
{\al _1\over z(\s _7)-z(\s_8)}-{\beta _1\over z(\s _7)-z(\s_8)}+...
\ee
So
\[{\p \tG \over \p x_1}=0 + \; higher \; order \;in \; \xn\]
We do not get any contribution.

\noindent
{\bf(iii)}
\be
 e^{\int \ko (\s _5 ).\ko (\s _6)[ \tG ]}\ki (\si ).\ko(\st ){\p
\tS\over
\p x_1(\si )} +
\si \leftrightarrow \st + \tS \leftrightarrow \tG
\ee
Using
${\p \tG \over \p x_1}=0$ the leading order contribution is zero.

Thus combining (i),(ii) and (iii)
 we conclude that there are no corrections to $\p _\mu F^{\mu
\nu}=0$ to this order.

\subsection{$\ki \ki , \kt$ Contributions to $\ytm$}

We start with, as usual, $e^A$ given by \eA

Pick out the terms that contribute to $\ytm$ involving $\ki \ki$ and
$\kt$.

There are four terms:

\noindent
{\bf (i)}
\be
 e^{\int \ko (\s _3 ).\ko (\s _4)[ \tG ]}\{ \ki (\si ).\ko(\st ){\p
\tS\over
\p x_1(\si )} + \si \leftrightarrow \st \}i\int \ki Y_1 \e
\ee
{\bf (ii)}
\be
 e^{\int \ko (\s _3 ).\ko (\s _4)[ \tG ]}\{ \kt (\si ).\ko(\st ){\p
\tS\over
\p x_2(\si )} + \si \leftrightarrow \st \} \e
\ee
{\bf (iii)}
\be 
e^{\int \ko (\s _3 ).\ko (\s _4)[ \tG ]}\{ \ki (\si ).\ki(\st )
\dsb \}
\e
\ee
{\bf (iv)}
\be
e^{\int \ko (\s _3 ).\ko (\s _4)[ \tG +\tS ]}i\int \kt Y_2 \e
\ee

We expand $\tS$ using (\ref{Approx}) and vary w.r.t. $\SI$:

\noindent
{\bf (i)}
\[
 e^{\int \ko (\s _3 ).\ko (\s _4)[ \tG ]}\{ \ki (\si ).\ko(\st ){\p
\bar \SI \over
\p x_1} + \si \leftrightarrow \st \}i\int \ki Y_1 \e
\]
\[
=
 e^{\int \ko (\s _3 ).\ko (\s _4)[ \tG ]} \ki (\si ).\ko(\st ){\p
\SI \over
\p x_1} i\int \ki (\s ) Y_1 \e
\]

${\delta \over \delta \SI}$ gives:
\be     \label{45}
- e^{\int \ko (\s _3 ).\ko (\s _4)[ \tG ]} \ki (\si ).\ko(\st )
 i\int \ki (\s ) Y_2 \e
\ee

We have to make contractions of the $\kn$'s. Let $z(\si )= z$ and
assign the momentum $p$ to this point. Let $z(\s )=w$ and assign
momentum $q$. Now $\st = \si$ or $\st =\s$ are two possibilities and
for each of these we can have $\s _3 =\si \; , \s _4 = \s $ or $\s _3
= \s \; , \s _4 = \si$.  None of the above need regularization. We can
also include the following two possibilities that need a regulator:$\s
_3 = \s _4 = \s$ or $\s _3 = \s _4 = \si$. For these cases we will let
$\s _A$ and $\s _B$ be the ``point splitting'' of $\s$ . Thus ($\s _3 = \s _A$
and $\s _4 = \s _B$) or ($\s _3 = \s _B$ and $\s _4 = \s _A$). We will
weight these with a factor of 1/2. This gives $q^2 \; ln \;
\epsilon$. Similarly point splitting $\si$ gives $p^2 \; ln \;
\epsilon$. Putting all the above together and using (\ref{A1})we get:
\be     \label{46}
-A(p).(p+q)iA^\mu (q) |z-w|^{2p.q}(\epsilon )^{p^2+q^2}Y_2^\mu e^{i(p+q)Y}
\ee
If we multiply and divide by $\epsilon ^{2p.q}$ this becomes
\be
-A(p).(p+q)iA^\mu (q) |{z-w\over \epsilon}|^{2p.q}(\epsilon )^{(p+q)^2}Y_2^\mu e^{i(p+q)Y}
\ee
Interchanging the role of $\s$ and $\si$ i.e. setting $z(\si )=w$ and
$z(\s )=z$,
 we get the same expression
with $p,q$ interchanged. Since these are just dummy variables 
we can combine the two if we allow both $p$ and $q$ to vary over the
full range of values.
\be
- A(p).(p+q)iA^\mu (q) |{z-w\over \epsilon}|^{2p.q}(\epsilon )^{(p+q)^2}Y_2^\mu e^{i(p+q)Y}
\ee
However 

There is also the possibility that $\s = \si$. In this case we point
split and let $\s = \s _A \; , \si = \s _B$ or vice versa (with weight
1/2 to each). Then using
 (\ref{S11}) we get 
\be     \label{49}
-S_{1,1}^{\mu \nu}\kon iY_2^{\mu}\e (\epsilon )^{\ko ^2}
\ee
{\bf (ii)}
\be
 e^{\int \ko (\s _5 ).\ko (\s _6)[ \tG ]}\{ \kt (\si ).\ko(\st ){\p
\tS\over
\p x_2(\si )} + \si \leftrightarrow \st \}\e
\ee

Using the approximations:
\[
 \dsix \tS (\si ,\st ) \approx
{\p \bar \SI \over \p x_2} = {1\over 2}{\p \SI \over \p x_2}
\]
\be     \label{Approx2} 
 \tS \approx \bar \SI = \SI 
\ee

\be
 e^{\int \ko (\s _5 ).\ko (\s _6)[ \tG ]} \kt (\si ).\ko(\st ){\p
\SI\over
\p x_2} \e
\ee
${\delta \over \delta \Sigma}$ gives
\be     \label{53}
-e^{\int \ko (\s _5 ).\ko (\s _6)[ \tG ]} \kt (\si ).\ko(\st )
i\kom Y_2^\mu \e
\ee
\be     \label{54}
=- (\epsilon )^{\ko ^2}S_2(\ko ).\ko i\kom Y_2 ^\mu
\ee
{\bf (iii)}
\be
 e^{\int \ko (\s _5 ).\ko (\s _6)[ \tG ]} \ki (\si ).\ki(\st )
{\pp \tS \over \p x_1(\si )\p x_1(\st )}
\ee
\be
{\pp \tS \over \p x_1(\si )\p x_1(\st )}
 \approx {1\over 2}({\pp \SI \over \p x_1^2 }- {\p \SI \over \p x_2})+...
\ee

\be
= e^{\int \ko (\s _5 ).\ko (\s _6)[ \tG ]}
 \ki (\si ).\ki(\st )
  {1\over 2}({\pp \SI \over \p x_1^2 }- {\p \SI \over \p x_2})
\e
\ee
${\delta \over \delta \SI}$
\be
= e^{\int \ko (\s _5 ).\ko (\s _6)[ \tG ]} \ki (\si ).\ki(\st )i\ko
Y_2 \e
\ee
Following the procedures described earlier this gives (when $\si \neq \st$):
\be
(\epsilon )^{(p+q)^2}|{z-w\over \epsilon }|^{2p.q}A(p).A(q)i(p+q)^\mu
Y_2^\mu
\e
\ee
Both $p$ and $q$ are integrated over the entire range.

When $\si =\st$ one has to point split and this gives:
\be
-(\epsilon ) ^{\ko ^2} i S_2 ^{\mu}(\ko )Y_2 ^\mu \e
\ee
{\bf (iv)}
\be
= e^{\int \ko (\s _5 ).\ko (\s _6)[\tS+ \tG ]} i\ktm Y_2^\mu
\ee
Using $\tS \approx \SI $ and varying w.r.t.$\SI$ gives
\be
= e^{\int \ko (\s _5 ).\ko (\s _6)[ \tG ]} \ko(\si ).\ko (\st ) i\ktm
Y_2^\mu \e
\ee
\be
=-(\epsilon )^{\ko ^2} \ko ^2 iS_2^\mu (\ko ) Y_2^\mu \e
\ee

We can also check that the equations are invariant at the loop
variable level:

\begin{eqnarray}
\delta (i)& = &
-\li (\s )\ko (\si ).\ko (\st )i\ki Y_2 \e e^{\int \ko (\s _5 ).\ko
(\s _6)
[ \tG ]}\\ &    &
-\li (\s )\ki (\si ).\ko (\st )i\ko Y_2 \e e^{\int \ko (\s _5 ).\ko
(\s _6)
[ \tG ]}\nonumber
\\
\delta (ii)& = &
-\li (\s )\ki (\si ).\ko (\st )i\ko Y_2 \e
 e^{\int \ko (\s _5 ).\ko (\s _6)[ \tG ]} \\
&    &-\lt (\s )\ko (\si ).\ko (\st )i\ko Y_2 \e
 e^{\int \ko (\s _5 ).\ko (\s _6)[ \tG ]}\nonumber
\\
\delta (iii)& =
&2\li (\s )\ki (\si ).\ko (\st )i\ko Y_2 \e 
 e^{\int \ko (\s _5 ).\ko (\s _6)[ \tG ]}\\
\delta (iv) & = &
\li (\s )\ko (\si ).\ko (\st )i\ki Y_2 \e
 e^{\int \ko (\s _5 ).\ko (\s _6)[ \tG ]} \\&    &
+\lt (\s )\ko (\si ).\ko (\st )i\ko Y_2 \e 
e^{\int \ko (\s _5 ).\ko (\s _6)[ \tG ]} \nonumber
\end{eqnarray}

Clearly the variations add up to zero. In Section 6 we will discuss
the gauge transformation law for space-time fields.

\subsection{$\ki \ki \ki , \kt \ki , k_3$ Contribution to $\yim$}

There are many terms that contribute. We will consider only the
 following term to
illustrate the technique being used.  

\noindent
{\bf (i)}
\[
{1\over 2!}
e^{\ko(\s _7).\ko (\s _8)\tG}
[\ki (\si ).\ko (\st ){\p \tS \over \p x_1 (\si )} + \si
\leftrightarrow \st ]\]\be
[\ki (\s _3 ).\ko (\s _4 ){\p \tG \over \p x_1 (\s _3 )} + \s _3
\leftrightarrow \s _4 ]
[\ki (\s _5 ).\ko (\s _6 ){\p \tG \over \p x_1 (\s _5 )} + \s _5
\leftrightarrow \s _6 ]\e =0.
\ee

We start with contractions that do not involve any regularization
(``non singular case'').
We will treat the cases that require regularization (``singular
case'') separately. In each case depending on the number of distinct
points we have different contributions.

{\bf Non Singular Case:}

\noindent
{\bf Three Vertex Operators:}

We first consider contractions involving three distinct verex
operators. Let us designate $\s _I , \s _{II}$ and $\s _{III}$ as 
the labels of the vertex operators with locations
$z(\s _I)=u , z(\s _{II})=w , z(\s _{III})=z$ and momenta $p,q$ and $ k$
 respectively
being associated with these vertex operators.

We use the approximation that $\tS \approx \SI$ as before and
integrate by parts on $x_1$ to get:
\[
{1\over 2!}
e^{\ko(\s _7).\ko (\s _8)\tG}
[\ki (\si ).\ko (\st ){\p \SI \over \p x_1 (\si )} + \si
\leftrightarrow \st ]\]\be
[\ki (\s _3 ).\ko (\s _4 ){\p \tG \over \p x_1 (\s _3 )} + \s _3
\leftrightarrow \s _4 ]
[\ki (\s _5 ).\ko (\s _6 ){\p \tG \over \p x_1 (\s _5 )} + \s _5
\leftrightarrow \s 6 ]\e
\ee
\[
=-
{1\over 2!}
e^{\ko(\s _7).\ko (\s _8)\tG}
\ki (\si ).\ko (\st )
[\ki (\s _3 ).\ko (\s _4 ){\p \tG \over \p x_1 (\s _3 )} + \s _3
\leftrightarrow \s _4 ]\]\be
[\ki (\s _5 ).\ko (\s _6 ){\p \tG \over \p x_1 (\s _5 )} + \s _5
\leftrightarrow \s _6 ]i\ko Y_1\e
\ee

Let us first consider contractions that do not involve
regularization. Thus consider the assignments
\br
p \leftrightarrow 
z(\s _I)= u & \si = \s _I      & \st = \s _I,\s _{II}, \s _{III} \nonumber \\
q \leftrightarrow  z(\s _{II})=w & \s _3 = \s _{II} & \s _4 = \s _I, \s _{III} \nonumber \\
k \leftrightarrow z(\s _{III})=z & \s _5= \s _{III} & \s _6= \s _I, \s _{II}
\er 
 This gives (using (\ref{A1})
\[
|{z -w\over \eps }|^{2k.q}|{w -u\over \eps }|^{2p.q}|{z -w\over \eps }|^{2p.k}
(\eps )^{(p+q+k)^2}\]
\be
A(p).(p+q+k)[{A(q).p\over w-u}+{A(q).k \over w-z}]
[{A(k).p\over z-u}+{A(k).q \over z-w}]
i(p+q+k)^\mu Y ^\mu _1 e^{i(p+q+k)}
\ee
{\bf Two Vertex Operators}

Now we come to the case where there are two vertex operators. The
assignment that does not need regularization is:

\noindent
{\bf (I)}
\br
p \leftrightarrow z(\s _I)=w & \si = \s _I & \st = \s _{I},\s
_{II}\nonumber \\
&\s _3 = \s _I & \s _4 = \s _{II} \nonumber \\
q \leftrightarrow z(\s _{II})=z &\s _5 = \s _{II}& \s _6 = \s _I
\er 

This gives:
\be
|{z-w\over \eps }|^{2p.q}(\eps )^{(p+q)^2}S_{1,1}^{\mu
\nu}(p)(p+q)^{\mu}q^\nu
{1\over w-z}A^\rho (q)p^\rho {1\over z-w}i(p+q)^\s Y_1^\s e^{i(p+q)Y}
\ee

The other possible assignment is:

\noindent
{\bf (II)}
\br
p \leftrightarrow z(\s _I)=w & \si = \s _I & \st = \s _{I},\s
_{II}\nonumber \\
q \leftrightarrow z(\s _{II})=z 
&\s _3 = \s _{II} & \s _4 = \s _{I} \nonumber \\
&\s _5 = \s _{II}& \s _6 = \s _I
\er 

which gives
\be
|{z-w\over \eps }|^{2p.q}(\eps )^{(p+q)^2}A^{\mu}(p)(p+q)^{\mu}
S_{1,1}^{\nu \rho}(q) {p^\nu p^\rho \over
 (w-z)^2}i(p+q)^\s Y_1^\s e^{i(p+q)Y}
\ee
{\bf Singular Cases:}

\noindent
{\bf Three Vertex Operators:}

We now consider assignments that require regularization. For the three
vertex operator case we have:
\br
p \leftrightarrow 
z(\s _I)= u & \si = \s _I      & \st = \s _I,\s _{II}, \s _{III} \nonumber \\
q \leftrightarrow  z(\s _{II})=w & \s _3 = \s _{II_A} & \s _4 = \s
_{II_B}, \nonumber \\
k \leftrightarrow z(\s _{III})=z & \s _5= \s _{III} & \s _6= \s _I, \s _{II}
\er

\[
|{z -w\over \eps }|^{2k.q}|{w -u\over \eps }|^{2p.q}|{z -w\over \eps }|^{2p.k}
(\eps )^{(p+q+k)^2}
\ki (\s _A).[\ko (\s _I)+\ko (\s _{II})+\ko(\s _{III})]\]
\be
\underbrace{
[{\ki (\s_A).\ko (\s _B )\over w_A-w_B} +{\ki (\s_B).\ko (\s _A )\over w_B-w_A}]}_{=0} 
([{\ki (\s _{III}.\ko (\s_I)\over z-u} +{\ki (\s _{III}.\ko (\s_I)\over z-w}] 
\ee

We have used $\s _A = \s _{II_A}$ and $\s _B = \s _{II_B}$. In the
limit $\s _A \rightarrow \s _B$, $<\ki (\s _A)>=<\ki  (\s _B)>= A(q)$
and 
$<\ko (\s _A)>=<\ko  (\s _B)>= q$
which is why the expression in square brackets vanishes.

\noindent
{\bf Two Vertex Operators:}

We turn to the case with two vertex operators.

\noindent
{\bf (I)} 
\br     \label{I}
p \leftrightarrow z(\s _I)=w & \si = \s _{I_A} & \st = \s _{I},\s
_{II}\nonumber \\
&\s _3 = \s _{I_B} & \s _4 = \s _{I_A} \nonumber \\
q \leftrightarrow z(\s _{II})=z &\s _5 = \s _{II}& \s _6 = \s _I
\er 
(We also have to consider the assignment with $\s _A$ and $\s _B$
interchanged.)

This gives:
\be     \label{Ia}
\ki (\s _A).[\ko (\s _A) + \ko (\s _{II})][{\ki (\s _B).\ko (\s
_A)\over w_B-w_A} (+ \s _3 \leftrightarrow \s _4 \; is \; not \;
allowed)]
[{\ki(\s _{II}).\ko(\s _I)\over z-w}]
\ee

(As before we are using $\s _{A, B}$ to denote $\s _{I_{A,B}}$.).

$\s _3 $ cannot be set to $\s _A$ because $\si = \s _A$. This is why
the exchange term involving $\s _3 $ and $\s _4$ is not allowed.   

If we interchange $A \leftrightarrow B$ in (\ref{Ia}) we find 
\be     \label{Ib}
\ki (\s _B).[\ko (\s _B) + \ko (\s _{II})][{\ki (\s _A).\ko (\s
_B)\over w_A-w_B} (+ \s _3 \leftrightarrow \s _4 \; is \; not \;
allowed)]
[{\ki(\s _{II}).\ko(\s _I)\over z-w}]
\ee
In the limit $\s _A \rightarrow \s _B$,  

$<\kim (\s _A) \kin (\s _B)> = S_{1,1}^{\mu \nu}$
and $\ko (\s _A)= \ko (\s _B) = \ko (\s _I)=p$
and thus (\ref{Ia}) and (\ref{Ib}) add up to zero.

\noindent
{\bf II}

\br
p \leftrightarrow z(\s _I)=w & \si = \s _I & \st = \s _{I},\s
_{II}\nonumber \\
q \leftrightarrow z(\s _{II})=z 
&\s _3 = \s _{II_A} & \s _4 = \s _{II_B} \nonumber \\
&\s _5 = \s _{II_B}& \s _6 = \s _{II_A}
\er

This gives:
\[
\ki (\s _I).[\ko (\s _I ) + \ko (\s _{II})][{\ki (\s _A).\ko (\s
_B)\over z_A-z_B} {\ki(\s _B).\ko (\s _A)\over z_B-z_A}]
i(p+q)^\mu Y_1^\mu e^{i(p+q)Y}
\]
\be
+\ki (\s _I).[\ko (\s _I ) + \ko (\s _{II})][{\ki (\s _B).\ko (\s
_A)\over z_B-z_A} {\ki(\s _A).\ko (\s _B)\over z_A-z_B}]
i(p+q)^\mu Y_1^\mu e^{i(p+q)Y}
\ee
We have used the same shorthand notation as in previous examples.
Note that just as in previous cases interchanging $\s _3$ and $\s _4 $
is not allowed. Note also that the two terms do not cancel. They add
to give:

\be
-A(p).(p+q){S_{1,1}^{\mu \nu}(q)q^\mu q ^\nu\over
\eps ^2}
\ee
We have weighted it by a factor $1\over 2$ as in (\ref{..}).

The final result is
\be
-A(p).(p+q){S_{1,1}^{\mu \nu}(q)q^\mu q ^\nu \over
\eps ^2}|{z-w\over \eps }|^{2p.q}(\eps )^{(p+q)^2}
\ee
Integrals over $w$ and $p,q$ are implicit.

\noindent
{\bf One Vertex Operator}

Finally we have to consider the assignment where there is only one
vertex operator and this clearly is singular and needs
regularization. We will also observe a serious dependence on the
prescription. This is not necessarily unacceptable. Presumably
different prescriptions involve field redefinitions. If we impose
physical state conditions on the fields these dependences should
disappear. 

\br
 z(\s _A )=z_A & \si = \s _A & \st = \s \nonumber \\
z(\s _B)=z_B & \s _3 = \s _B & \s _4 = \s _A ,\s _C \nonumber \\
z(\s _C) = z_C & \s _5 = \s _C & \s _6 = \s _A, \s _B 
\er
We assign the momentum $k$ to the vertex operator.
There are a total of 3! ways of assigning labels. So we weight each
 possibility by $1\over 3!$. What is given above is only one of the
 possibilities.

It gives
\[
\ki (\s _A).[\ko (\s _A) + \ko (\s _B) + \ko (\s _C)][
{\ki (\s _B).\ko (\s _A)\over z_B-z_A}+{\ki (\s _B).\ko (\s _C)\over z_B-z_C}]
\]\be
[{\ki (\s _C).\ko (\s _B)\over z_C-z_B}+
{\ki (\s _C).\ko (\s _A)\over z_C-z_A}] 
\ee
Now we take the three points to be equidistant (this is a
prescription) and this implies that
\br
z_B - z_A & = & \eps \nonumber \\
z_C - z_B & = & \eps \nonumber \\
z_C - z_A & = & 2\eps 
\er

The expression in the second pair of 
square brackets vanishes and this term is zero.
 The term obtained by interchanging $\s _A $ and $\s _C$ also
vanishes. Similarly the two terms that have $\s _5 = \s _B$ also
vanish. Thus four of the six possibilities give zero. The remaining
two are given by the assignment
\br
 z(\s _A )=z_A & \si = \s _B & \st = \s \nonumber \\
z(\s _B)=z_B & \s _3 = \s _A & \s _4 = \s _A ,\s _C \nonumber \\
z(\s _C) = z_C & \s _5 = \s _C & \s _6 = \s _A, \s _B 
\er
and the one obtained by interchanging $\s _A$ and $\s _C$ in this.
This gives:
\[
\ki (\s _B).[\ko (\s _A) + \ko (\s _B) + \ko (\s _C)][
{\ki (\s _A).\ko (\s _B)\over z_A-z_B}+{\ki (\s _A).\ko (\s _C)\over z_A-z_C}]
\]\be
[{\ki (\s _C).\ko (\s _A)\over z_C-z_A}+
{\ki (\s _C).\ko (\s _B)\over z_C-z_B}] 
\ee

Plugging the space time fields and the rest of the factors we get
\be
S_{1,1,1}^{\mu \nu \rho}(\ko ) \kom 
[{\kon \over (-\eps )}+{\kon \over (-2\eps )}]
[{\ko ^\rho \over (2\eps )} +{\ko ^\rho \over (\eps )}] (\eps
)^{\ko^2}
i\ko ^\s Y_1^\s  
\ee
\be
= -{3\over 4}S_{1,1,1}^{\mu \nu \rho }\kom \kon \ko ^\rho (\eps )^{\ko ^2 -2}i\ko
^\s Y_1^\s
\ee
We have multiplied the answer by $2\over 3!$ as the weight for this
term.

We have thus calculated the contribution from the first term to the
the equation of motion.

 What is to be noted is that the field $S_{1,1,1}$ is present
 as a result of the fact that we did not throw away the singular
 (normal ordering) pieces. This term will be indispensable in defining
 gauge transformations because there will be terms that cannot be
 assigned to any other field - in fact the presence of this term
 therefore guarantees that a gauge transformation can
 always be defined.

\section{Space-Time Fields and their Transformations}
\setcounter{equation}{00}

Now we proceed to define fields. In the first approximation they were
defined \cite{BS2,Puri} by the following equations:
\br
<\kim > & = & A_1^\mu \nonumber \\
<\kim (\si ) \kin (\st )> & = & {D(\si - \st )\over a}S^{\mu
\nu}_{1,1}+ A_1^\mu A_1^\nu \nonumber \\
<\ktm > & = & S_2^\mu
\er

We will define the gauge transformation laws for the space time fields
by comparing the variations of the loop variable expression with the
field expression. Thus consider expression (i)in Section 5.4 in both
forms:
\vskip5mm

\noindent
{\bf A(i)} given in (\ref{46}) and (\ref{49}):

\be
- A(p).(p+q)iA^\mu (q) |{z-w\over \epsilon}|^{2p.q}(\epsilon
)^{(p+q)^2}Y_2^\mu e^{i(p+q)Y} -S_{1,1}^{\mu \nu}\kon iY_2^{\mu}\e 
(\epsilon )^{\ko ^2}
\ee
{\bf B(i)} given in (\ref{45}):
\be
- e^{\int \ko (\s _5 ).\ko (\s _6)[ \tG ]} \ki (\si ).\ko(\st )
 i \ki (\s ) Y_2 \e
\ee
Integrals over $z,w$ and $p,q,k$ are implicit in all the above. Thus
the integral 
\be 
\int dz \int dw |{z-w\over \epsilon}|^{2p.q} = \int dz F(p,q)
\ee

$F(p,q)$ is defined only after suitable regularization. The actual
evaluation of this function will be done later. 

Now we consider the variation of B(i):
\br
\delta B(i) & = & [-i\li (\s ) \ko (\si ).\ko (\st ) \kim (\s _3 )
Y_2^\mu  \\& & -i\li (\s ) \ki (\si ).\ko (\st ) i \kom (\s
_3 )Y_2 ^\mu]
e^{\ko (\s _5).\ko (\s_6)\tG}\e \nonumber
\er 
We convert this to space -time fields :
\br
\delta B(i) &= &[-i\Lambda ^\mu _{1,1}(k )\ko ^2 (\epsilon )^{\ko
^2} e^{i\ko Y} Y_2^\mu \nonumber \\ & &-i\Lambda _1 (p) A_1^\mu (q) 
(p+q)^2|{z-w\over \epsilon}|^{2p.q}
\epsilon ^{(p+q)^2}e^{i(p+q)Y}Y_2^\mu ]\nonumber \\
&  & +[-i\Lambda _{1,1}^\nu (k)\kon \kom (\epsilon )^{\ko ^2} e^{i\ko Y}
Y_2^\mu  \\ & & 
-i \Lambda _1 (q) A_1^\nu (p) (p+q)^\nu (p+q)^\mu |{z-w\over \epsilon
}|^{2p.q}(\epsilon )^{(p+q)^2} e^{i(p+q)}Y_2^\mu ]\nonumber
\er
This is to be compared with $\delta A(i)$:
We will write 
\br
\delta S^{\mu \nu}_{1,1} (k)e^{i\ko Y}&=& [\Lambda _{1,1}^\mu (k)\kon +
\Lambda _{1,1}^\nu (k) \kom + \delta _{int} S^{\mu \nu}_{1,1}] \nonumber \\
\delta A_1^\mu (p)e^{i p_0 Y} & = &p^\mu \Lambda _1(p)e^{ip_0 Y} 
\er 
and determine $\delta _{int} S_{1,1}^{\mu \nu}$

We get
\[
\delta B(i)  =  \delta _{free} A(i)+ \]\[
 +\int dp dq \delta (p+q-k)[-iq.(p+q)\Lambda _1(p)A_1^\mu (q)
-i\Lambda _1 (q)p^\mu A(p).(p+q)]F(p,q)(\epsilon
)^{(p+q)^2}e^{i(p+q)Y}+\]
\be
 =\delta _{free}A(i) + 
\delta _{int} S_{1,1}^{\mu \nu}e^{i\ko Y}(\epsilon )^{\ko ^2}\kon 
\ee
This fixes $\delta _{int} S_{1,1}^{\mu \nu}$ to be
 
\be     \label{dS}
\delta _{int}S_{1,1}^{\mu \nu}(k) =
\int dp dq \delta (p+q-k)[-i\Lambda _1(p)q^\nu A_1^\mu (q)
-i\Lambda _1 (q)p^\mu A _1(p)^\nu ]F(p,q)
\ee
{\bf B(ii)} (\ref{53})
\[
e^{\ko (\s _5).\ko (\s_6)\tG}\kt (\si ).\ko (\st )i\kom (\s _3 ) e^{i\ko
Y}Y_2^\mu 
\]
 
\[
\delta B(ii) =
- e^{\ko (\s _5).\ko (\s_6)\tG}[\li (\s )\ki (\si ).\ko (\st )
+ \lt (\s ) \ko (\si ).\ko (\st )]i\kom (\s ) e^{i\ko
Y}Y_2^\mu 
\]
\[
=
- \Lambda _{1,1}^{\mu}(\ko )\kon i\kom (\eps )^{\ko ^2}e^{i\ko Y}
\]
\[-\int dp dq \delta (p+q-k)\int dw \Lambda _1(p)A_1^\nu (q)
(p_0 +q_0)^\nu i (p_0 +q_0)^\mu |{z-w\over
\eps}|^{2p.q}(\eps ) ^{(p+q)^2}e^{i(p+q)Y}Y_2^\mu
\]
\be     \label{Bii} 
-\Lambda _2 (\ko )\ko ^2 i\kom (\eps )^{\ko ^2}Y_2^\mu e^{i\ko Y}
\ee
{\bf A(ii)} (\ref{54})
\[
A(ii)=-S_2^\nu (\kon ) i\ko (\eps )^{\ko ^2}Y_2 ^\mu e^{i\ko Y}
\]
\be
\delta S_2^\mu (\ko )= \Lambda _2 (\ko )\kom + \Lambda _{1,1}^\mu (\ko
) + \delta _{int}S_2^\mu
\ee
\[
\delta A(ii)=
 -\Lambda _2 (\ko )\ko ^2 i\kom (\eps )^{\ko ^2}Y_2^\mu e^{i\ko Y}
\]
\[
- \Lambda _{1,1}^{\mu}(\ko )\kon i\kom (\eps )^{\ko ^2}e^{i\ko Y}
\]
\be     \label{Aii}
 - \delta _{int} S_2 ^\mu \kon i\kom (\eps ) ^{\ko ^2}Y_2^\mu e^{i\ko Y}
\ee
Comparing (\ref{Bii}) with (\ref{Aii}) we find
\[
\delta B(ii)= \delta A(ii) +  \delta S_{2\; int}^\mu \kon i\kom (\eps
) ^{\ko ^2}Y_2^\mu e^{i\ko Y}- \]
\be \int dpdq \delta (p+q-k) \Lambda _1(p)A_1^\nu (q)(p_0 +q_0)^\nu
i (p_0 +q_0)^\mu \underbrace{\int dw |{z-w\over
\eps}|^{2p.q}}_{F(p,q)}
(\eps ) ^{(p+q)^2}e^{i(p+q)Y}Y_2^\mu
\ee
From this we conclude that
\be
\delta _{int}S_2^\mu (\ko ) =  \int dp dq \delta (p+q-k)\Lambda
_1(p)A_1^\nu (q)F(p,q)
\ee
Thus we obtain the transformation rules for $S_{1,1}$ and
$S_2$. Equations (iii) and (iv) are clearly consistent with this since
they differ only in index structure.

It is not particularly illuminating to describe in detail the gauge
transformation law for $S_{1,1,1}$ that one obtains in this manner
since the calculation is very similar to that of $S_{1,1}$.

\section{Consistency of Gauge Transformations and $\xn$-dependence of
Fields}
\setcounter{equation}{00}

We examine, in this section, the question of {\em consistency} of
gauge transformations of space-time fields defined in earlier sections.
The question arises because there are different equations that can be
used to define the gauge transformation law of $S_{1,1}$. For instance
when one integrates by parts on $x_1$, different vertex operators
such as $Y_1$ or $Y_2$ are obtained depending on whether one
differentiates $e^{i\ko Y}$ twice or acts once each on $e^{i\ko Y}$
and $e^{\ko .\ko \tG}$. The dependence on $z-w$ is thus different and
one obtains instead of $F(p,q)$ (in (\ref{dS})) some other function, and
thus a different transformation law.

In fact $F(p,q)$ is a function of $\xn$ because $\tG (z-w) = ln \, (z-w)
+ O(\xn )$.  Thus in principle one can ask what the result of
differentiating (\ref{dS}) by $\xn$ is. The RHS of (\ref{dS}) is
non-zero on differentiating and one reaches an
inconsistency unless one assumes that the LHS also is non-zero - i.e.
it must be a function of $\xn$ as well. This leads inexorably to the
conclusion that the space-time fields such as $S_{1,1}$ must be
functions of $\xn$, $S_{1,1}(\ko , \xn )$. 

In the equation defining $S$ (\ref{S11}) there is a natural way to
introduce this dependence, and this is to make the ``string field''
$\Psi$ a function of $\xn$. Thus:
\be     \label{xn}
\int d k_n d\lambda _n \kim \kin \Psi [k_n, \ko , \la _n , \xn ]=
S_{1,1}^{\mu \nu}(\ko , \xn )
\ee

Since  $\la _n$ is to lowest order the shift in $\xn$, we can change
variables \cite{BS1} to $y_n$, defined by \[\sum _n \la _n t^{-n}=e^{\sum _n
t^{-n}y_n}\] and replace (\ref{xn}) by
\be     
\int d k_n dy_n \kim \kin \Psi [k_n, \ko , y_n , \xn ]=
S_{1,1}^{\mu \nu}(\ko , \xn )
\ee

Both $\xn$ and $y_n$ are gauge coordinates. It is necessary therefore
to understand the presence of both of these coordinates in the $\Psi$.
Our starting point is a field $\Psi (X(z), \xn +y_n )$. The breakup of
the gauge coordinate into $\xn +y_n$ is similar in spirit to that in
the background field method in field theory. We treat $\xn$ as a
background or reference point. Now we do a generalized Fourier
transform using the loop variable and define the variable $Y$ which
has in it only $\xn$. Thus the relation between the original string
field and the one we have been using in this paper can be
summarized in the following way: (We use the symbol $\Psi$ for all the
fields - the arguments of the fields will
make clear which field we are referring to)
\[
\Psi (X, \xn +y_n) = \Psi (\yn ,\xn ,y_n)= 
\int [d\kn ] \Psi (\kn , \xn , y_n) \gvk \]

Thus while the original field is only a function of $\xn +y_n$, once
we define the variable $Y$ we have specified a reference point. The
space-time fields obtained by (\ref{S11}) thus depend on this
reference point. Gauge invariance is the statement that physics is
independent of $\xn +y_n$. In terms of the new variables it becomes
independence of $\xn$. Thus the $\kn ,y_n$ integrals are in the nature
of Fourier transformations, whereas the $\xn$ integral is an
imposition of gauge invariance.
  
One can also do the
integral over $\ko$:
\[ \int d\ko S_{1,1}(\ko , \xn )e^{i\ko Y}= S_{1,1}(Y,\xn ). \] 
Note $S$ depends explicitly on $\xn$ but also implicitly on $\xn$
through $Y$ because $Y$ depends on $\xn$ 
and all the derivatives of $X(z)$. Thus
$S$ is a non-local object in that it depends on all the derivatives of
$X$. To put it another way, specifying $S$ requires specifying a curve
$X(z)$, because no two curves will produce the same value for $S$ for
all $\xn$. 

Thus the dependence on the infinite number of $\xn$
coordinates effectively makes $S$ non-local in $z$ and therefore
$X(z)$. Of course the relevant scale here is the string scale so at
low energies one can neglect the higher derivatives and effectively
$S$ becomes an ordinary local field. It is also possible to redefine
fields so that this non-locality disappears \cite{BSP}.

Note also that $\xn$ being along gauge directions we can fix gauge and
set them to some fixed value. So there is no increase in the number of
physical degrees of freedom. This is a desirable feature.

\section{Conclusions}

We have described a general construction that gives gauge
invariant equations of motion, the gauge transformation prescription
(in terms of loop variables) being the same as in II. This method was
outlined in III but many of the details had not been worked out.
One of the problems that was left unsolved was whether the map from
loop variables to space time fields is unambiguous. In particular it
was not obvious that there was a map that correctly reproduced the
gauge transformations. The results of the present paper indicate that
it is indeed possible to define space time fields and their gauge
transformations consistently. There are two crucial ingredients. One
is that one has to carefully keep all the singular terms that are
normally discarded by ``normal ordering''. We have to keep a finite
cutoff in order for this procedure to make sense. This is not unexpected - we
already know that in order to define
off-shell Green functions in this approach, one needs a finite cutoff 
\cite{BSPT,BSOS,AD}.
As shown in \cite{BSFC}, even U(1) gauge invariance of the massless
vector is violated when a finite cutoff is introduced in order to go  
off-shell, and one needs to introduce massive modes to restore gauge  
invariance.  In the loop variable formalism all the modes
are present from the start and there is no problem.
Gauge invariance is present, on or off-shell.
However the exact value of the Koba-Nielsen integral will depend
on the cutoff prescription. Presumably these are equivalent to field 
redefinitions (of the space time fields).  

 The second ingredient is that the string-field $\Psi$ and
thus the space-time fields are functions of the gauge coordinates $\xn$.
This is crucial for consistency of the definitions of gauge
transformations. Thus effectively ``space-time'' has become infinite
dimensional! 

At this stage we have a non-trivial interacting theory with an
infinite tower of higher spin gauge fields and a large gauge
invariance.  By construction these modes are essentially 
\footnote{``essentially'' because of a technicality that is discussed
in I} those of the 
open bosonic
string (including the auxiliary ones). 
Nevertheless we have not proved that the
amplitudes of this theory are those of the bosonic string. We have to
demonstrate that the procedure of ``dimensional reduction'' that
worked for the free case goes through here also i.e. without loss to
gauge invariance. If this works out
then we are guaranteed that the on-shell amplitudes
are those governed by the bosonic string simply because the two
dimensional correlators that are being calculated here are identical
to those of the bosonic string amplitude calculation.
There are arguments that this is in fact the case \cite{BSP}. Furthermore as
the gauge invariance does not use any on-shell conditions, these
amplitudes are guaranteed to be gauge invariant off-shell also. Thus
we have an off shell formulation. Further tests of the consistency of
this will involve checking loop amplitudes. This is work for the
future.

The main 
advantages are that the prescription for writing
down the equations and gauge transformation laws are fairly
straightforward.  The gauge transformations written in terms of
loop variables seem to have some geometric meaning - they look
like local scale transformations.  The interactions look
as if they have the effect of converting a string to a membrane.
The fields also appear massless in one higher dimension.
These are intriguing features.
\cite{SZ,BP,W}
 Finally, assuming the above issues are 
resolved satisfactorily, one has to see whether this formalism 
provides any insight into the various other issues that have become
pressing in string theory, such as duality. Some of the structure
observed in \cite{BSV} may be relevant for this.

\appendix

\renewcommand{\thesection}{\Alph{section}}
\renewcommand{\theequation}{\thesection.\arabic{equation}}

\section{Appendix: Covariant Taylor Expansion}
\label{appena}
\setcounter{equation}{0}

We derive the covariant Taylor expansion for $\tS (z(\si), z(\st ),\si
,\st )$. We first derive a Taylor expansion for $Y(z)$ and then use it
to obtain a Taylor expansion for $\tS$.
\subsection{Taylor Expansion for Y}
Ordinary Taylor expansion gives,
\be
Y(z+a) = Y(z) + a {dY\over dz} + {a^2\over 2}{d^2Y\over dz^2} +...
\ee
\be     \label{Y}
Y \equiv \sum _{n\ge 0}\aln \tY _n
\ee 
where $\tY _n \equiv {1\over (n-1)!}{\p ^n X \over \p z^n}$ and $\aln$
satisfy \cite{BS1}
 \[
\sum _{n\ge 0} \aln t^{-n}= e^{\sum _{n\ge 0}\xn t^{-n}}
\]
\[
{\p \aln \over \p \xm }= \al _{n-m} \; ,
\]
\[
{dY\over dz}= \tY _1 + \sum _{n=1}n \aln \tY_{n+1} \; ,
\]
\be     \label{Y'}
=\tY _1 + \sum _{n=1}n\xn {\p \over \p x_{n+1}}Y \;.
\ee
In the above we have used $\sum _n [ n\xn {\p \over \p \xn }]\al _m = m
\al _m $.

Differentiating (\ref{Y'}) gives
\[ {d^2 Y\over dz^2}=
 \tY _2 +
 \sum _{n=1}n\xn {\p \over \p x_{n+1}}
({dY\over dz})\]
Plugging in (\ref{Y'})
\[
=\tY _2 + \sum _{n=1}n\xn {\p \over \p x_{n+1}}(\tY _1 + \sum _{m=1}m\xm
{\p Y\over \p x_{m+1}})\]
\be     \label{Y''}
{d^2 Y\over dz^2}= \tY _2 +  \sum _{n,m=1}nm \xn \xm  {\p Y\over \p x_{n+m+2}}
+ \sum _{m=2}m(m-1)x _{m-1} {\p Y \over \p x_{m+1}}
\ee
Adding (\ref{Y}),(\ref{Y'}),(\ref{Y''}) gives the first few terms of a
Taylor series, except that we would like to express $\tY _i$ in terms
of $Y_n$ in order to make the expression covariant.

{\bf $\tY$ in terms of Y:} 

We first write
\be
\al (t) \p _z X(z+t) = \sum _{n,m\ge 0}t^{m-n}\aln \tY _{m+1}
\ee
Let $\beta (t) = \sum _{p\ge 0} \beta _p t^{-p}$. Let us evaluate
$\int {dt\over t}\beta (t) \al (t) \p _z X(z+t)$.
\[
\int {dt \over t}\beta (t) \al (t) \p _z X(z+t) 
= \sum _{n,p,m=n+p}\beta _p \aln \tY
_{m+1}
\]
\[
= \sum _{m,p \ge 0}
\beta _p {\p \over \p x_{p+1}}
\al _{m+1}\tY_{m+1}\]
So
\be     \label{beta}
\int {dt \over t}\beta (t) \al (t) \p _z X(z+t) 
=
\sum _{p\ge 0}\beta _p {\p \over \p x_{p+1}}Y
\ee

Let us choose $\beta (t)$ having the property $\beta (t) \al (t) =
t^{-s} ; s \ge 0$ and call it $\beta ^s (t)$ with the expansion 
\[ \beta ^s (t) = \sum _{p\ge s}\beta ^s _p t^{-p}\]  
Then (\ref{beta}) will become
\be
\tY _{s+1}= \sum _{p\ge s}\beta ^s _p Y_{p+1}
\ee

Thus if we determine $\beta ^s _p$ we obtain the required expansion.

To determine $\beta ^s _p$ we note that 
\[ \beta ^s _p t^{-p}= \beta ^s(t) = t^{-s} \al ^{-1} (t)\]
\[=t^{-s}e^{-\sum _n \xn t^{-n}}\]
\[= \sum _{n\ge 0}\aln (-\xn )t^{-n-s}= \sum _{p\ge s}
\al _{p-s}(-\xn ) t^{-p}\]
\[={\p \over \p (-x_s)}\al _p (-\xn ,t)\]
This gives 
\be
{\p \al _p \over \p x_s}\mid _{\xn \rightarrow -\xn}=\al _{p-s}(-\xn )=
\beta ^s _p
\ee

Thus for instance 
\[ \beta ^0_0=1\]\[ \beta ^0_1=-x_1 \]\[ \beta ^0_2={x_1^2\over
2}-x_2\]
\[ \beta ^0_3=-{x_1^3\over 6}+x_2x_1 -x_3 \]
 Therefore
\be
\tY_1=Y_1 -x_1Y_2 +({x_1^2\over 2}-x_2 )Y_3 + ...
\ee

It is easy to see that \[ \beta ^s _p = \beta ^0_{p-s}\] and so all
the coefficients are easily determined.

Similarly
\[{\p \over \p \xn }\beta ^s _r = - \beta ^s _{r-n}\] Using this it is
easy to see that
\[ \dsn \tY _s =0\] as it should be.

Using the above results one obtains
\[
{dY\over dz} =\tY _1 + \sum _{n=1}n\xn {\p \over \p \xn}Y_1
\]
\[
=\sum _{r\ge 0}(\beta ^0_r + r x_r )Y_{r+1}
\]
\be
=\sum _{r\ge 0}\gamma ^0_r Y_{r+1}
\ee 

Using the fact that $\dsnm Y = \dsq Y$ one can easily verify that $\dsnm
{dY\over dz}= \dsq {dY\over dz}$. This is as it should be 
because the operations of
differentiating w.r.t. z and w.r.t $\xn$ commute.

\subsection{Taylor Expansion of $\tS$}
We now use this to obtain the covariant expansion of $\tS$.
$\tS$ was defined in section III
\be
\tS (z_1,z_2) = \oint du \omega  (u)<\p _u X (u) Y(v-a)><\p _u X(u) Y'(v+a)>
\ee
where $z_2-z_1 =2a$ and $z_1 +z_2=2v$ and the contour encircles both points.
We will call the gauge coordinates $\xn$ at $z_1$ and $y_n$ at
$z_2$. The prime on $Y$ indicates indicates that it is a function
of $y_n$.

We will use the shorthand notation $<Y(v-a)Y'(v+a)>$ for the above
definition of $\tS$.
Thus we have 
the following Taylor expansion for $\tS$:
\[<Y(v-a)Y'(v+a)>= <(Y(v)-a {dY\over dv} +
 {a^2\over 2}{d^2Y\over dv^2}+...)\]
\[(Y'(v)-a{dY'\over dv}+ 
 {a^2\over 2}{d^2Y'\over dv^2}+...)>\]
Plugging in the Taylor expansions for $Y$ that has been derived in
this Appendix we get 

\[=\bar \SI (v) + a\sum _{r\ge 0}(\gamma _r^{0'}{\p \over \p y_{r+1}} - 
\gamma _r^0 {\p \over \p x_{r+1}} )\bar \SI (v)+\] 
\be
a^2 [-\sum _{r,s\ge 0}\gamma _r^0 \gamma _s^{0'}{\p ^2 \over \p
x_{r+1}\p y_{s+1}}\bar \SI + {1\over 2}
\sum _{r\ge 0} (\gamma _r^1{\p \over \p x_{r+1}}  + 
 \gamma _r^{'1}{\p \over \p y_{r+1}})\bar \SI ]  + O(a ^3)
\ee 
Here as before $\bar \SI (z,\xn ,y_n) \equiv \tS (z,\xn ,z,y_n)$

This is what has been used in section VI.

\end{document}